\documentclass{eptcs} 
\providecommand{\event}{QPL 2017} % Name of the event you are submitting to
\usepackage{underscore}           % Only needed if you use pdflatex.
\usepackage{amsfonts,amsmath,amssymb,amsthm}
\usepackage[usenames]{color}

\newtheorem{theorem}{Theorem}%[section]

\newtheorem{corollary}[theorem]{Corollary}
\newtheorem{definition}[theorem]{Definition}

\newtheorem{proposition}[theorem]{Proposition}
\newtheorem{remark}[theorem]{Remark}

\newcommand{\scr}[1]{\ensuremath{\mathcal {#1}}}

\newcommand{\eps}{\varepsilon}

\newcommand{\notarrow}{\kern .42em\not\kern -.42em\longrightarrow}
\newcommand{\ket}[1]{\ensuremath{|#1\rangle}}
\newcommand{\bra}[1]{\ensuremath{\langle#1|}}
\renewcommand{\phi}{\varphi}

\newcommand{\nm}[1]{\ensuremath{\Vert #1 \Vert}}

\renewcommand{\H}{\mathcal H}
\newcommand{\K}{\mathcal K}
\renewcommand{\L}{\mathcal L}

\renewcommand{\O}{\mathcal O}
\newcommand{\R}{\mathbb R}

\newcommand{\V}{\mathcal V}

\newcommand{\Dim}[1]{\text{Dim}(#1)}
\newcommand{\Tr}[1]{\text{Tr}(#1)}

\newcommand{\noprint}[1]{\relax}

\title{Common Denominator for Value and Expectation No-go Theorems:
Extended Abstract}
\author{Andreas Blass
\institute{University of Michigan}
\and
Yuri Gurevich
\institute{Microsoft and University of Michigan}
}
\def\titlerunning{Common Denominator}
\def\authorrunning{Blass and Gurevich}
\begin{document}
\maketitle

\begin{abstract}
Hidden-variable (HV) theories allege that a quantum state describes an ensemble of systems distinguished by the values of hidden variables. No-go theorems assert that HV theories cannot match the predictions of quantum theory. The present work started with repairing ﬂaws in the literature on no-go theorems asserting that HV theories cannot predict the expectation values of measurements. That literature gives one an impression that expectation no-go theorems subsume the time-honored no-go theorems asserting that HV theories cannot predict the possible values of measurements. But the two approaches speak about different kinds of measurement. This hinders comparing them to each other. Only projection measurements are common to both. Here, we sharpen the results of both approaches so that only projection measurements are used. This allows us to clarify the similarities and differences between the two approaches. Neither one dominates the other. 
\end{abstract}

\section{Introduction}

Hidden-variable theories allege that a state of a quantum system, even if it is pure and thus contains as much information as quantum mechanics permits, actually describes an ensemble of systems with distinct values of some hidden variables. Once the values of these variables are specified, the system becomes determinate or at least more determinate than quantum mechanics says.  Thus the randomness in
quantum predictions results, entirely or partially, from the randomness involved in selecting a member of the ensemble. No-go theorems assert that, under reasonable hypotheses, a hidden-variable interpretation cannot reproduce the predictions of quantum mechanics. 

In this paper, we examine two species of such theorems, \emph{value} and \emph{expectation} no-go theorems.
The value approach originated in the work of Bell \cite{bell64,bell66} and of Kochen and Specker \cite{ks} in the 1960's. Value no-go theorems establish that, under suitable hypotheses, hidden-variable theories cannot reproduce the predictions of quantum mechanics of the possible values of observables. 
The expectation approach was developed by Spekkens \cite{spekkens} and by Ferrie, Emerson, and Morris \cite{ferrieA,ferrieB,ferrieC}, with \cite{ferrieC} giving the sharpest result.  Expectation no-go theorems establish that, under suitable hypotheses, hidden-variable theories cannot reproduce the predictions of quantum mechanics of the expectations of observables. 

The bold philosophical paper \cite{spekkens} of Spekkens attracted our attention. We started \cite{G226,G228} with repairing various flaws in the four papers on expectation no-go theorems. In \cite[\S VI.B]{ferrieB}, Ferrie and Emerson write that ``Spekkens'
notion of non-contextuality is a generalization of the traditional notion initiated by Kochen and Specker.'' This and other remarks strongly suggest that the expectation approach is more general than the value approach. We had our doubts about that.

In both the value and the expectation approach, measurements are associated to Hermitian operators, but they are different sorts of measurements.  In the value approach, Hermitian operators serve as observables, and measuring one of them
produces a number in its spectrum.  In the expectation approach, certain Hermitian operators serve as effects. Here an effect is a positive operator $E$ dominated by the identity operator $I$ (so that $I-E$ is positive as well). Measuring an effect $E$ produces 0 or 1, even if the spectrum of $E$ consists entirely of other numbers. The only Hermitian operators for which these two uses coincide are
projections.

We sharpen the results of both approaches so that only projection
measurements are used. Regarding the expectation approach, we
substantially weaken the hypotheses. Arbitrary effects are replaced with rank-1 projections. Accordingly, we need convex-linearity
only for the hidden-variable picture of states, not for that of
effects. Regarding the value approach, it turns out that rank-1
projections are sufficient in the finite dimensional case but not in
general. Finally, using a successful hidden-variable theory of John
Bell for a single qubit, we demonstrate that the expectation approach does not subsume the value approach.

A more comprehensive version of this paper, complete with the proofs, is found at \cite{G234}.

\section{Expectation No-Go Theorem}
\label{sec:exp}

\begin{definition}\label{def:exp}\rm
An \emph{expectation representation} for quantum systems described by a Hilbert space $\H$ is a triple $(\Lambda,\mu,F)$ where
\begin{itemize}
\item $\Lambda$ is a measurable space,
\item $\mu$ is a convex-linear map assigning to each density operator $\rho$ on $\H$ a probability measure $\mu(\rho)$ on $\Lambda$, and
\item $F$ is a map assigning to each rank-1 projection $E$ in $\H$ a measurable function $F(E)$ from $\Lambda$ to the real interval $[0,1]$.
\end{itemize}
It is required that for all density matrices $\rho$ and all rank-1 projections $E$
\begin{equation}\label{eq1}
\Tr{\rho\cdot E}=\int_\Lambda F(E)\,d\mu(\rho)
\end{equation}
\end{definition}

\bigskip
The convex linearity of $\mu$ means that $\mu(p_1\rho_1 + p_2\rho_2) = p_1\mu(\rho_1) + p_2\mu(\rho_2)$ whenever $p_1,p_2$ are nonnegative real numbers with sum 1.

The definition of expectation representation is similar to  Ferrie-Morris-Emerson's definition of the probability representation \cite{ferrieC} except that (i)~the domain of $F$ contains only rank-1 projections, rather than arbitrary effects, and (ii)~we do not (and cannot) require that $F$ be convex-linear.

Intuitively an expectation representation $(\Lambda,\mu,F)$ attempts to predict the expectation value of any rank-1 projection $E$ in a given mixed state $\rho$. The hidden variables are combined into one variable ranging over $\Lambda$. Further, $\mu(\rho)$ is the probability measure on $\Lambda$ determined by $\rho$, and $(F(E))(\lambda)$ is the probability of determining the effect $E$ at the subensemble of $\rho$ determined by $\lambda$. The left side of \eqref{eq1} is the expectation of $E$ in state $\rho$ predicted by quantum mechanics and the right side is the expectation of $F(E)$ in the ensemble described by $\mu(\rho)$.

But why is $\mu$ supposed to be convex linear? Well, mixed states have
physical meaning and so it is desirable that $\mu$ be defined on mixed
states as well. If you are a hidden-variable theorist, it is most
natural for you to think of a mixed state as a classical probabilistic
combination of the component states. This leads you to the convex
linearity of $\mu$. For example, if $\rho = \sum_{i=1}^k p_i\rho_i$
where $p_i$'s are nonnegative reals and $\sum p_i = 1$ then, by the
rules of probability theory, $(\mu(\rho))(S) = \sum p_i(\mu(\rho_i))(S)$
for any measurable $S\subseteq\Lambda$. Note, however, that you cannot
start with any wild probability distribution $\mu$ on pure states and
then extend it to mixed states by convex linearity. There is an
important constraint on $\mu$. The same mixed state $\rho$ may have
different representations as a convex combination of pure states; all
such representations must lead to the same probability measure
$\mu(\rho)$.

\medskip
\begin{theorem}[First Bootstrapping Theorem]\label{thm:boot1}
Let $\H$ be a closed subspace of a Hilbert space $\H'$.  From any expectation representation for quantum systems described by $\H'$, one can directly construct such a representation for systems described by $\H$.
\end{theorem}

\medskip
\begin{theorem}[Expectation no-go theorem]\label{thm:exp}
  If the dimension of the Hilbert space $\H$ is at least 2 then there is no expectation representation for quantum systems described by $\H$.
\end{theorem}

We cannot expect any sort of no-go result in lower dimensions, because quantum theory in Hilbert spaces of dimensions 0 and 1 is trivial and therefore classical.
By the First Bootstrapping Theorem, it suffices to prove Theorem~\ref{thm:exp} just in the case $\Dim{\H}=2$.
But we find Ferrie-Morris-Emerson's proof that works directly for
all dimensions \cite{ferrieC} instructive, and in the full paper \cite{G234} we adjust it to prove
Theorem~\ref{thm:exp}.
The adjustment involves adding some details and observing that a
drastically reduced domain of $F$ suffices. The adjustment also
involves making a little correction. Ferrie et al. quoted an erroneous
result of Bugajski \cite{bugajski} which needs some additional
hypotheses to become correct. Fortunately for Ferrie et al., those
hypotheses hold in their situation.

\begin{remark}[Symmetry or the lack of thereof]\rm
In view of the idea of symmetry or even-handedness suggested by
Spekkens \cite{spekkens}, one might ask whether there is a dual
version of Theorem~\ref{thm:exp}, that is, a version that requires
convex-linearity for effects but looks only at pure states and does not
require any convex-linearity for states.
The answer is no; with such requirements there is a trivial example of a successful hidden-variable theory, regardless of the dimension of the Hilbert space.  The theory can be concisely described as taking the quantum state itself as the ``hidden'' variable.  In more detail, let $\Lambda$ be the set of all pure states.  Let $\mu$ assign to each operator $\ket\psi\bra\psi$ the probability measure on $\Lambda$ concentrated at the point $\lambda_{\ket\psi}$ that corresponds to the vector $\ket\psi$.  Let $F$ assign to each effect $E$ the function on $\Lambda$ defined by $F(E)(\lambda_{\ket\psi})=\bra\psi E\ket\psi$.
We have trivially arranged for this to give the correct expectation
for any effect $E$ and any pure state \ket\psi.  The formula for
$F(E)$ is clearly convex-linear (in fact, linear) as a function of
$E$.  Of course, $\mu$ cannot be extended convex-linearly to mixed
states, so that Theorem~\ref{thm:exp} does not apply.
\end{remark}

\section{Value No-Go Theorems}
\label{sec:val}
Value no-go theorems assert that hidden-variable theories cannot even produce the correct outcomes for individual measurements, let alone the correct probabilities or expectation values.  Such theorems considerably predated the expectation no-go theorems considered in the preceding section.  Value no-go theorems were first established by Bell \cite{bell64,bell66} and then by Kochen and Specker \cite{ks}; we shall also refer to the user-friendly exposition given by Mermin \cite{mermin}.  To formulate value no-go theorems, one must specify what ``correct outcomes for individual measurements'' means.

\begin{definition} \label{valmap}
Let $\H$ be a Hilbert space, and let $\O$ be a set of observables, i.e., self-adjoint operators on $\H$. A \emph{valuation} for $\O$ in $\H$ is a function $v$ assigning to each observable $A\in\O$ a number $v(A)$ in the spectrum of $A$, in such a way that $(v(A_1),\dots,v(A_n))$ is in the joint spectrum $\sigma(A_1,\dots,A_n)$ of $(A_1,\dots,A_n)$
whenever $A_1,\dots,A_n$ are pairwise commuting.
\end{definition}

The intention behind this definition is that, in a hidden-variable theory, a quantum state represents an ensemble of individual systems, each of which has definite values for observables. That is, each individual system has a valuation associated to it, describing what values would be obtained if we were to measure observable properties of the system.  A believer in such a hidden-variable theory would expect a valuation for the set of all self-adjoint operators on $\H$, unless there were superselection rules rendering some such operators unobservable.

Before we proceed, we recall the notion of joint spectra \cite[Section~6.5]{spectral}.

\medskip
\begin{definition}
The \emph{joint spectrum} $\sigma(A_1,\dots,A_n)$ of pairwise
commuting, self-adjoint operators $A_1,\dots,A_n$ on a Hilbert space
$\H$ is a subset of $\R^n$. If $A_1,\dots,A_n$ are simultaneously
diagonalizable then $(\lambda_1,\dots,\lambda_n)
\in\sigma(A_1,\dots,A_n)$ iff there is a non-zero vector $\ket\psi$
with $A_i\ket\psi=\lambda_i\ket\psi$ for $i=1,\dots,n$. In general,
$(\lambda_1,\dots,\lambda_n) \in\sigma(A_1,\dots,A_n)$ iff for every
$\eps>0$ there is a unit vector $\ket\psi\in\H$ with
$\nm{A_i\ket\psi-\lambda_i\ket\psi}<\eps$ for $i=1,\dots,n$.
\end{definition}

\begin{proposition}\label{pro:jspec}\
For any continuous function $f:\R^n\to\R$, we have\\ 
$f(A_1,\dots,A_n)=0$ if and only if $f$ vanishes identically on $\sigma(A_1,\dots,A_n)$.
\end{proposition}

The proposition is implicit in the statement, on page~155 of \cite{spectral}, that ``most of Section~1, Subsection~4, about functions of a one operator,'' can be repeated in the context of several commuting operators.  We give a detailed proof of the proposition in \cite[\S4.1]{G228}.

\begin{theorem}[\cite{bell66,ks,mermin}]\label{thm:dim3}
If $\Dim{\H}=3$ then there is a finite set $\O$ of rank~1 projections for which no valuation exists.
\end{theorem}

The proof of Theorem~\ref{thm:dim3} can be derived from the work of
Bell \cite[Section~5]{bell66}, and we do that explicitly in
\cite[\S4.3]{G228}. The construction given by Kochen and Specker
\cite{ks}  provides the desired $\O$ more directly.  The proof of
Theorem~1 in \cite{ks} uses a Boolean algebra generated by a finite
set of one-dimensional subspaces of $\H$, and it shows that the
projections to those subspaces constitute an $\O$ of the required
sort.  Mermin's elegant exposition \cite[Section~IV]{mermin} deals
instead with squares $S_i^2$ of certain spin-components of a spin-1
particle, but these are projections to 2-dimensional subspaces of
$\H$, and the complementary rank-1 projections $I-S_i^2$ serve as the
desired $\O$.

\begin{theorem}[Second Bootstrapping Theorem]\label{thm:boot2}
Suppose $\H\subseteq\H'$ are finite-dimensional Hilbert spaces.  Suppose further that $\O$ is a finite set of rank-1 projections of $\H$ for which no valuation exists.  Then there is a finite set $\O'$ of rank-1 projections of $\H'$ for which no valuation exists.
\end{theorem}

Intuitively, such dimension bootstrapping results are to be expected. If hidden-variable theories could explain the behavior of quantum systems described by the larger Hilbert space, say $\H'$, then they could also provide an explanation for systems described by the subspace $\H$.  The latter systems are, after all, just a special case of the former, consisting of the pure states that happen to lie in $\H$ or mixtures of such states. But often no-go theorems give much more information than just the impossibility of matching the predictions of quantum-mechanics with a hidden-variable theory.  They establish that hidden-variable theories must fail in very specific ways. It is not so obvious that these specific sorts of failures, once established for a Hilbert space $\H$, necessarily also apply to its
superspaces $\H'$.

\begin{theorem}[Value no-go theorem]\label{thm:val}
Suppose that the dimension of the Hilbert space is at least 3.
\begin{enumerate}
\item There is a finite set $\O$ of projections for which no valuation exists.
\item If the dimension is finite then there is a finite set $\O$
  of rank~1 projections for which no valuation exists.
\end{enumerate}
\end{theorem}

The desired finite sets of projections are constructed explicitly in the proof \cite{G234}. The finiteness assumption in part (2) of the theorem cannot be omitted. If $\Dim{\H}$ is infinite, then the set $\O$ of all finite-rank projections admits a valuation, namely the constant zero function.  This works because the definition of ``valuation'' imposes constraints on only finitely many observables at a time.

Let's say that a projection $A$ on Hilbert space $\H$ is a \emph{rank-$n$ projection modulo identity} if either $A$ is of rank $n$ or else $\H$ splits into a tensor product $\K\otimes\L$ such that $\K$ is finite-dimensional and $A$ has the form $P\otimes I_{\L}$ where $P$ is of rank $n$ and $I_{\L}$ is the identity operator on $\L$. The proof of Theorem~\ref{thm:val} gives us the following corollary.

\begin{corollary}
If the dimension of the Hilbert space is at least 3 then there is a finite set of rank-1 projections modulo identity for which no valuation exists.
\end{corollary}

\section{One successful hidden-variable theory}
\label{sec:bell}

By reducing both species of no-go theorems to projection measurement,
where measurement as observable and measurement as effect coincide, we
made it easier to see similarities and differences.  No, the
expectation no-go theorem does not imply the value no-go theorem. But
the task of proving this claim formally, say for a given dimension
$d=\Dim{\H}$, is rather thankless. You have to construct a
counter-factual physical world where the expectation no-go theorem
holds but the value no-go theorem fails. There is, however, one
exceptional case, that of dimension~2. Theorem~\ref{thm:exp} assumes
$\Dim{\H}\ge2$ while Theorem~\ref{thm:val} assumes $\Dim{\H}\ge3$. So
what about dimension 2?

Bell developed, in \cite{bell64} and \cite{bell66}, a hidden-variable theory for a two-dimensional Hilbert space $\H$. Here we summarize the improved version of Bell's theory due to Mermin \cite{mermin}, we simplify part of Mermin's argument, and we explain why the theory doesn't contradict Theorem~\ref{thm:exp}.

In the rest of this section, we work in the two-dimensional Hilbert space $\H$. Let $\V$ be the set of value maps $v$ for all the observables on $\H$ (so that $v(A)$ is an eigenvalue of an observable $A$).  In each pure state $\ket\psi$, the hidden variables should determine a particular member of $\V$.

\begin{definition}\label{def:val}\rm
A \emph{value representation} for quantum systems described by $\H$ is a pair $(\Lambda,V)$ where
\begin{itemize}
\item $\Lambda$ is a probability space and
\item $V$ a function $\ket\psi\to V_{\ket\psi}$ on the pure states such that every $V_{\ket\psi}$ is a map $\lambda\to V_{\ket\psi}^\lambda$ from $\Lambda$ into $\V$.
\end{itemize}
Further, we require that, for any pure state $\ket\psi$ and any observable $A$, the expectation $\int_\Lambda V_{\ket\psi}^\lambda(A)\: d\lambda$ of the eigenvalue of $A$ agrees with the prediction $\bra{\psi}A\ket{\psi}$ of quantum theory:
\begin{equation}\label{bell1}
 \int_\Lambda V_{\ket\psi}^\lambda(A)\: d\lambda = \bra{\psi}A\ket{\psi}
\end{equation}
\end{definition}

\medskip
Definition~\ref{def:val} is narrowly tailored for our goals in this section; for a general definition of value representation see \cite{G228}. Notice that, if a random variable (in our case, the eigenvalue of $A$ in $\psi$) takes only two values, then the expected value determines the probability distribution. A priori we should be speaking about commuting operators and joint spectra but things trivialize in the 2-dimensional case. Recall Proposition~\ref{pro:jspec} and notice that, in the 2-dimensional Hilbert space, if operators $A,B$ commute, then one of them is a polynomial function of the other.

\medskip
\begin{theorem}\label{thm:bell}
There exists a value representation for the quantum systems described by the two-dimensional Hilbert space $\H$.
\end{theorem}

A natural question arises why Bell's theory doesn't contradict Theorem~\ref{thm:exp}? This is related to the convex  linearity requirement. To obtain an expectation representation, we must extend the value representation, constructed in the proof of Theorem~\ref{thm:bell}, to all density matrices in a convex linear way. But no such extension exists.  In the full paper \cite{G234}, we give a simple example illustrating what goes wrong if one attempts to extend the value representation in a convex linear way.
Thus, Bell's example of a hidden-variable theory for 2-dimensional
\scr H does not fit the assumptions in any of the expectation no-go
theorems.  It does not, therefore, clash with the fact that those
theorems, unlike the value no-go theorems, apply in the 2-dimensional
case.

\newpage
\nocite{*}
\bibliographystyle{eptcs}
\bibliography{nogo}

\begin{thebibliography}{10}
\providecommand{\bibitemdeclare}[2]{}
\providecommand{\surnamestart}{}
\providecommand{\surnameend}{}
\providecommand{\urlprefix}{Available at }
\providecommand{\url}[1]{\texttt{#1}}
\providecommand{\href}[2]{\texttt{#2}}
\providecommand{\urlalt}[2]{\href{#1}{#2}}
\providecommand{\doi}[1]{doi:\urlalt{http://dx.doi.org/#1}{#1}}
\providecommand{\bibinfo}[2]{#2}

\bibitemdeclare{article}{bell64}
\bibitem{bell64}
\bibinfo{author}{John~S. \surnamestart Bell\surnameend} (\bibinfo{year}{1964}):
  \emph{\bibinfo{title}{On the Einstein-Podolsky-Rosen paradox}}.
\newblock {\sl \bibinfo{journal}{Physics}} \bibinfo{volume}{1}, pp.
  \bibinfo{pages}{195--200},
  \doi{10.1103/PhysicsPhysiqueFizika.1.195}.

\bibitemdeclare{article}{bell66}
\bibitem{bell66}
\bibinfo{author}{John~S. \surnamestart Bell\surnameend} (\bibinfo{year}{1966}):
  \emph{\bibinfo{title}{On the problem of hidden variables in quantum
  mechanics}}.
\newblock {\sl \bibinfo{journal}{Reviews of Modern Physics}}
  \bibinfo{volume}{38}, pp. \bibinfo{pages}{447--452},
  \doi{10.1103/RevModPhys.38.447}.

\bibitemdeclare{book}{spectral}
\bibitem{spectral}
\bibinfo{author}{Michael~S. \surnamestart Birman\surnameend} \&
  \bibinfo{author}{Michael~Z. \surnamestart Solomjak\surnameend}
  (\bibinfo{year}{1987}): \emph{\bibinfo{title}{Spectral Theory of Self-Adjoint
  Operators in Hilbert Space}}.
\newblock \bibinfo{publisher}{Reidel}.
\newblock \bibinfo{note}{Originally in Russian, Leningrad University Press,
  1980}.

\bibitemdeclare{article}{G234}
\bibitem{G234}
\bibinfo{author}{Andreas \surnamestart Blass\surnameend} \&
  \bibinfo{author}{Yuri \surnamestart Gurevich\surnameend} (\bibinfo{year}{July
  2017}): \emph{\bibinfo{title}{Common denominator for value and expectation
  no-go theorems}}.
\newblock \bibinfo{note}{\urlalt{https://arxiv.org/abs/1707.07368}{arXiv:1707.07368}}.

\bibitemdeclare{article}{G226}
\bibitem{G226}
\bibinfo{author}{Andreas \surnamestart Blass\surnameend} \&
  \bibinfo{author}{Yuri \surnamestart Gurevich\surnameend}
  (\bibinfo{year}{March 2015}): \emph{\bibinfo{title}{Spekkens's Symmetric
  No-Go Theorem}}.
\newblock \bibinfo{note}{\urlalt{https://arxiv.org/abs/1503.08084}{arXiv:1503.08084}}.

\bibitemdeclare{article}{G228}
\bibitem{G228}
\bibinfo{author}{Andreas \surnamestart Blass\surnameend} \&
  \bibinfo{author}{Yuri \surnamestart Gurevich\surnameend}
  (\bibinfo{year}{Sept. 2015}): \emph{\bibinfo{title}{On Hidden Variables:
  Value and Expectation No-Go Theorems}}.
\newblock \bibinfo{note}{\urlalt{https://arxiv.org/abs/1509.06896}{arXiv:1509.06896}}.

\bibitemdeclare{article}{bugajski}
\bibitem{bugajski}
\bibinfo{author}{S\l \surnamestart awomir Bugajski\surnameend}
  (\bibinfo{year}{1993}): \emph{\bibinfo{title}{Classical frames for a quantum
  theory---a bird's-eye view}}.
\newblock {\sl \bibinfo{journal}{International Journal of Theoretical Physics}}
  \bibinfo{volume}{32}, pp. \bibinfo{pages}{969--977},
  \doi{10.1007/BF01215303}.

\bibitemdeclare{article}{ferrieA}
\bibitem{ferrieA}
\bibinfo{author}{Christopher \surnamestart Ferrie\surnameend} \&
  \bibinfo{author}{Joseph \surnamestart Emerson\surnameend}
  (\bibinfo{year}{2008}): \emph{\bibinfo{title}{Frame representations of
  quantum mechanics and the necessity of negativity in quasi-probability
  representations}}.
\newblock {\sl \bibinfo{journal}{Journal of Physics A: Mathematical and
  Theoretical}} \bibinfo{volume}{41}:\bibinfo{eid}{352001},
  \doi{10.1088/1751-8113/41/35/352001}.
\newblock \bibinfo{note}{See also \urlalt{https://arxiv.org/abs/0711.2658}{arXiv:0711.2658}}.

\bibitemdeclare{article}{ferrieB}
\bibitem{ferrieB}
\bibinfo{author}{Christopher \surnamestart Ferrie\surnameend} \&
  \bibinfo{author}{Joseph \surnamestart Emerson\surnameend}
  (\bibinfo{year}{2009}): \emph{\bibinfo{title}{Framed Hilbert space: hanging
  the quasi-probability pictures of quantum theory}}.
\newblock {\sl \bibinfo{journal}{New Journal of Physics}}
  \bibinfo{volume}{11}:\bibinfo{eid}{063040},
  \doi{10.1088/1367-2630/11/6/063040}.
\newblock \bibinfo{note}{See also \urlalt{https://arxiv.org/abs/0903.4843}{arXiv:0903.4843}}.

\bibitemdeclare{article}{ferrieC}
\bibitem{ferrieC}
\bibinfo{author}{Christopher \surnamestart Ferrie\surnameend},
  \bibinfo{author}{Ryan \surnamestart Morris\surnameend} \&
  \bibinfo{author}{Joseph \surnamestart Emerson\surnameend}
  (\bibinfo{year}{2010}): \emph{\bibinfo{title}{Necessity of negativity in
  quantum theory}}.
\newblock {\sl \bibinfo{journal}{Physical Review A}}
  \bibinfo{volume}{82}:\bibinfo{eid}{044103}, \doi{10.1103/PhysRev.40.749}.
\newblock \bibinfo{note}{See also \urlalt{https://arxiv.org/abs/0910.3198}{arXiv:0910.3198}}.

\bibitemdeclare{article}{heino}
\bibitem{heino}
\bibinfo{author}{Teiko \surnamestart Heinosaari\surnameend}
  (\bibinfo{year}{2013}): \emph{\bibinfo{title}{A simple sufficient condition
  for the coexistence of quantum effects}}.
\newblock {\sl \bibinfo{journal}{Journal of Physics A: Mathematical and
  Theoretical}} \bibinfo{volume}{46}:\bibinfo{eid}{152002},
  \doi{10.1088/1751-8113/46/15/152002}.

\bibitemdeclare{article}{ks}
\bibitem{ks}
\bibinfo{author}{Simon \surnamestart Kochen\surnameend} \&
  \bibinfo{author}{Ernst \surnamestart Specker\surnameend}
  (\bibinfo{year}{1967}): \emph{\bibinfo{title}{The problem of hidden variables
  in quantum mechanics}}.
\newblock {\sl \bibinfo{journal}{Journal of Mathematics and Mechanics}}
  \bibinfo{volume}{17}, pp. \bibinfo{pages}{59--87}.

\bibitemdeclare{article}{mermin}
\bibitem{mermin}
\bibinfo{author}{N.~David \surnamestart Mermin\surnameend}
  (\bibinfo{year}{1993}): \emph{\bibinfo{title}{Hidden variables and the two
  theorems of John Bell}}.
\newblock {\sl \bibinfo{journal}{Reviews of Modern Physics}}
  \bibinfo{volume}{65}, pp. \bibinfo{pages}{803--815},
  \doi{10.1103/RevModPhys.65.803}.

\bibitemdeclare{book}{schatten}
\bibitem{schatten}
\bibinfo{author}{Robert \surnamestart Schatten\surnameend}
  (\bibinfo{year}{1970}): \emph{\bibinfo{title}{Norm ideals of completely
  continuous operators}}.
\newblock \bibinfo{publisher}{Springer Verlag},
  \doi{10.1007/978-3-662-35155-0}.
\newblock \bibinfo{note}{2nd edition}.

\bibitemdeclare{article}{spekkens}
\bibitem{spekkens}
\bibinfo{author}{Robert~W. \surnamestart Spekkens\surnameend}
  (\bibinfo{year}{2008}): \emph{\bibinfo{title}{Negativity and contextuality
  are equivalent notions of nonclassicality}}.
\newblock {\sl \bibinfo{journal}{Physics Review Lettters}}
  \bibinfo{volume}{101(2)}:\bibinfo{eid}{020401},
  \doi{10.1103/PhysRevLett.101.020401}.
\newblock \bibinfo{note}{See also \urlalt{https://arxiv.org/abs/0710.5549}{arXiv:0710.5549}}.

\end{thebibliography}
\end{document}